\documentclass[12pt]{article}

\usepackage{cite}
\usepackage{graphicx}
\usepackage{amssymb}
\usepackage{amsfonts}
\usepackage{amsmath}

\usepackage{hyperref}
\usepackage{fullpage}

\def\M2Pl{M^2_{\mathrm{Pl}}}
\def\MPl{M_{\mathrm{Pl}}}

\begin{document}

\title{Quartet-metric gravity and scalar graviton\\ dark holes: supplements}

\author{Yu.\ F.\ Pirogov\footnote{Yury.Pirogov@ihep.ru}\, and\,
        O.\ V.\ Zenin\footnote{Oleg.Zenin@ihep.ru}\\
        {\it NRC ``Kurchatov Institute'' -- IHEP, Protvino, Russia}}
\date{}
\maketitle


\begin{abstract}
This note includes results of a study of stationary spherically symmetric  ``dark holes'',
objects merging central black holes and  peripheral scalar graviton dark haloes 
arising in the framework of the modified gravity -- 
the quartet-metric, or more appropriately, multiscalar-metric gravity with spontaneously broken relativity as a gauge symmetry~\cite{Pirogov2015}.
An exact solution to be considered as a basic one for the dark holes is presented.
This solution may, in principle, be used for a qualitative description of 
the effect of asymptotically flat rotation curves in galaxies.
To convert the basic solution into a more realistic one suitable for astrophysical applications,
a family of modified solutions is considered.
These solutions are studied numerically for representative set of free parameters.

These results were reported in the talk~\cite{icppa-2024-talk} but not included 
into the proceedings~\cite{icppa-2024-proc} due to article length limitations.
The note can be considered as a supplement to this reference.
\end{abstract}

\maketitle

\section{Introduction} \label{sec:introduction}

The existence of (cold) dark matter (DM) is strongly motivated by cosmological and astrophysical observational data analyzed 
in the framework of General Relativity (GR) and the Standard Model of particle physics (SM)~\cite{PDG2024}.
On the other hand, the DM phenomenon raises a number of theoretical and experimental issues.
The SM contains no DM candidate, and there is no preferred path of SM extension providing it.
Experimental searches for DM particles did not yield any positive signal yet.
Cold DM, irrespective of its particular nature, exhibits dynamical problems at scales $\le$1~Mpc~\cite{PDG2024}. 

Given the DM problems emerging in the framework of GR and (extended) SM, one can alternatively search for ``dark'' degrees of freedom in a modified gravity.
GR, as an effective field theory (EFT) with the gauge symmetry group of diffeomorphisms (Diff), leaves only two on-shell degrees of freedom for the massless tensor graviton.
Additional gravitational degrees of freedom may originate from a spontaneous breaking of the Diff gauge symmetry, 
with extra polarizations of now massive tensor graviton and a physical scalar gravi-Higgs boson (scalar graviton) serving as dark energy (DE) and DM candidates.
We consider an implementation of the spontaneous breaking of the modified Diff gauge symmetry,
the multiscalar-metric gravity with spontaneously broken relativity (SBR)~\cite{Pirogov2015,Pirogov2017,Pirogov2021}, 
where the emerging physical gravi-Higgs boson/scalar graviton becomes a natural DM component~\cite{Pirogov2011}.

\section{The multiscalar-metric gravity with spontaneously broken relativity}

In the SBR framework, gravity is described by an EFT constructed in observer's arbitrary kinematic coordinates $x^\alpha$ on a dynamical metric $g_{\mu\nu}(x^\alpha)$ 
and a set of distinct dynamical coordinates $z^{\alpha} \equiv \delta^\alpha_a Z^a(x_\mu)$ where scalar fields $Z^a$, $a = 0,1,2,3$ 
are transformed (piecewise in $x^\alpha$) under the constant Poincar\'e group acting in $Z^a$ space.\footnote{%
Generally, the spacetime dimension $D$ is arbitrary.
For definiteness, here and in what follows $D=4$ is chosen 
(henceforth, the term quartet-metric gravity).
}
More specifically, the gravity theory is considered as a gauge theory corresponding to a spontaneosly broken relativity symmetry, with $Z^a$ being gravi-Higgs fields.
From particle physics viewpoint, 3 combinations of $Z^{a}$ components become additional components of a massive tensor graviton, while 
the remaining one combination of $Z^a$ describes a scalar graviton playing the r\^ole of a physical gravi-Higgs boson.

Poincar\'e symmetry in $Z$ space implies that $Z^a$ enter the Lagrangian only via a derivative term 
\begin{equation*}\label{eq:zeta-munu}
	\zeta_{\mu\nu} \equiv \partial_\mu Z^a \partial_\nu Z^b \eta_{ab} , 
\end{equation*}
where $\eta_{ab}$ is the Minkowski symbol, and the quasi-metric $\zeta_{\mu\nu}$  possesses (patchwise in $x^\alpha$) an inverse ${\zeta^{-1}}^{\mu\nu}$, with the determinant $\zeta \equiv \mathrm{det}(\zeta_{\mu\nu}) < 0$.
The fields $g_{\mu\nu}$ and $\zeta_{\mu\nu}$ enter the action in the form of effective fields. 
The effective scalar graviton field is defined as 
\begin{equation}\label{eq:sigma}
	\sigma = \log(\sqrt{-g}/\sqrt{-\zeta}) \, .
\end{equation}
The effective metric $\bar{g}_{\mu\nu}$ entering the observables and the metric/quasi-metric correlator $\bar{\ae}^\mu_\nu$ are defined, respectively, as
\begin{equation}\label{eq:gbar-ash}
	\bar{g}_{\mu\nu} \equiv e^{\bar{w}} g_{\mu\nu}, \, \bar{{\ae}}^\mu_\nu \equiv \bar{g}^{\mu\lambda} \zeta_{\lambda\nu} \, .
\end{equation}
Here $\bar{w} \equiv \bar{w}(\sigma)$ is the gravitational scale factor\footnote{Termed Weyl form-factor in \cite{icppa-2024-proc}.
E.g., $\bar w = -\sigma/2$  corresponds to a spontaneously broken Weyl-transverse relativity and $\bar w = 0$ to a spontaneosly broken GR.
}
and $\bar{g}_{\mu\nu}$ possesses an inverse ${\bar g}^{\mu\nu} \equiv {\bar g}^{-1\, \mu\nu}$.
$\bar{\ae}^\mu_\nu$ is a kind of a dynamical DE required to implement the mechanism of spontaneous breaking of the relativity symmetry~\cite{Pirogov2021}.
In terms of the effective fields, the action functional reads
\begin{equation}\label{eq:action-effective}
	S[g_{\mu\nu}, Z^a] = \int \bar{L}(\bar{g}_{\mu\nu}, \bar{{\ae}}^\mu_\nu, \sigma)\, \sqrt{-\bar{g}}\, d^4x \, ,
\end{equation}
where $\bar{L}$ defines a particular variant of the gravity model.

\paragraph*{The Lagrangian and field equations.}
Following Ref.~\cite{icppa-2024-proc}, we use the minimal Lagrangian, 
 \begin{equation} 
	\bar L   = \bar L_g + \bar L_s + \bar L_m \label{eq:L} \, ,
 \end{equation}
with the tensor and scalar parts defined, respectively, as 
\begin{equation}
	\bar L_g = -\frac{1}{2} \M2Pl\, \bar R(\bar g_{\mu\nu}) - \bar V_{{\ae}}({\bar{{\ae}}}^\mu_\nu) \label{eq:Ltensor} \, ,
\end{equation}
\begin{equation}
	\bar L_s =  \frac{1}{2} \bar g^{\mu\nu} \partial_\mu s \partial_\nu s - V_s(s) \label{eq:Lscalar} \,   .
\end{equation}
Here $\MPl = 1/(8\pi G_N)^{1/2}$ is the Planck mass and  $\bar R = R(\bar g_{\mu\nu})$ is the Ricci scalar curvature expressed via the effective metric.
The physical scalar graviton field is $s = M_s \sigma$, with the  scale $M_s < \MPl$.
The potential $\bar V_{{\ae}}$ depends on traces of products of $\bar{{\ae}}^{\mu}_{\nu}$ and (along with $V_s(s)$) is responsible for a spontaneous relativity breaking~\cite{Pirogov2021}.
As in Ref.~\cite{icppa-2024-proc}, we approximate $\bar V_{{\ae}}$ by an effective CC, $\bar V_{{\ae}} = \M2Pl\bar\Lambda$, and neglect the tensor graviton mass.
As long as pure vacuum configurations are considered, the ordinary matter term ${\bar L}_m$ is omitted.
${\bar L}_s$ can be conveniently expressed in terms of the dimensionless field $\sigma$, with the scale $M_s \equiv \MPl\Upsilon$ factored out:
\begin{equation*} \label{eq:Lscalar-sigma}
	{\bar L}_s = {\bar L}_\sigma = {\M2Pl\Upsilon^2} \left[ \frac{1}{2} \bar g^{\mu\nu} \partial_\mu \sigma \partial_\nu \sigma - V_\sigma(\sigma) \right] \, .
\end{equation*}
where $V_\sigma(\sigma) = V_s(\MPl\Upsilon\sigma)/\M2Pl\Upsilon^2$.
The dimensionless parameter $\Upsilon$ characterizes the coupling between scalar and tensor gravity.
Variation of action~(\ref{eq:action-effective}) with Lagrangian density~(\ref{eq:L}) with respect to the basic metric $g^{\mu\nu}$ gives
Einstein's equations in terms of the effective metric and $\sigma$ (with traceless terms explicitly separated):
\begin{eqnarray}
	\M2Pl \left[\bar R_{\mu\nu} -\frac{1}{4} \bar g_{\mu\nu} \bar R \right]  &=& %
	{\bar {T_\sigma}}_{\mu\nu} - \frac{1}{4} \bar g_{\mu\nu} {\bar T_\sigma} \, + \nonumber \\
	&+&   \frac{1}{4} \bar g_{\mu\nu} \left[ (1+2\bar w') \left( \M2Pl (\bar R  + 4\bar\Lambda) +  \bar T_\sigma\right)  - \frac{4}{\sqrt{- \bar g}} \frac{\delta \sqrt{- \bar g} \bar L_\sigma}{\delta\sigma} %
	\right] \, , \label{FE-tensor}
\end{eqnarray}
where
\begin{eqnarray} \label{FE-tensor-defs}
	{\bar R}_{\mu\nu} &=& {R}_{\mu\nu}(\bar g_{\mu\nu}) \nonumber \, ,\\
	{\bar {T_\sigma}}_{\mu\nu} &=& \M2Pl \Upsilon^2 \left[ \partial_\mu\sigma \partial_\nu\sigma %
	                               - \bar g_{\mu\nu} \left( \frac{1}{2} (\partial\sigma)^2 - V_\sigma \right)\right] \,, \,\,\,%
	\bar {T_\sigma} = \bar g^{\mu\nu} {\bar {T_\sigma}}_{\mu\nu} \label{eq:T-sigma}   \nonumber \, ,  \\
	\frac{\delta \sqrt{-\bar g} {\bar L}_\sigma}{\delta \sigma} &=& %
	- \M2Pl\Upsilon^2 \left[ \partial_\mu \left(\sqrt{-\bar g} \bar g^{\mu\nu} \partial_\nu \sigma \right)  %
	                         + \sqrt{- \bar g}\, \frac{\partial V_\sigma}{\partial\sigma} \right] \nonumber \, ,\\
	\bar w' &=& d\bar w(\sigma)/d\sigma \, \nonumber .
\end{eqnarray}
From Eq.~(\ref{FE-tensor}) one gets the constraint:
\begin{equation} \label{FE-tensor-trace}
	 \frac{1}{4} (1+2\bar w') \left( \M2Pl (\bar R + 4\bar\Lambda)  + \bar T_\sigma \right) %
		- \frac{1}{\sqrt{- \bar g}} \frac{\delta \sqrt{- \bar g} \bar L_\sigma}{\delta\sigma} = 0  \,\, .
\end{equation}
Variation of the action with respect to $Z^a$ gives
\begin{equation} \label{eq:FE-Z}
 	\partial_\nu \left[ \sqrt{-\bar g}\, {\zeta^{-1}}^{\mu\nu} \partial_\mu Z^a\, \frac{1}{4} \left( \M2Pl (\bar R + 4\bar\Lambda) + \bar T_\sigma \right) \right] = 0\, .
\end{equation}
This equation can be immediately integrated:\footnote{%
Without loss of generality, choose the coordinates $x^\mu \equiv z^\mu \equiv \delta^\mu_a Z^a$ where the equation takes a simplified form.
See Ref.~\cite{icppa-2024-proc} for details.
}
\begin{equation}\label{eq:FE-Z-integrated}
	e^{2\bar w +\sigma}\, \frac{1}{4}\left(\M2Pl (\bar R + 4\bar\Lambda) + {\bar T}_\sigma \right)  = - \M2Pl {\kappa_s} \, ,
\end{equation}
where $\M2Pl\kappa_s$ is an arbitrary integration constant.\footnote{$\kappa_s$ was denoted $\Lambda_0$ in \cite{Pirogov2011}.}

From Eqs. (\ref{FE-tensor}), (\ref{FE-tensor-trace}), and (\ref{eq:FE-Z-integrated}) one finally gets:
\begin{eqnarray} 
	{\bar R}_{\mu\nu} - \frac{1}{2} {\bar g}_{\mu\nu} \bar R &=& %
	 \frac{1}{\M2Pl} \bar T^{\mathrm{eff}}_{\mu\nu} + \bar g_{\mu\nu} \bar\Lambda %
	 \label{eq:Gmunu}  \, ,\\
	\frac{1}{\sqrt{-\bar g}}\, \partial_\mu \left(\sqrt{-\bar g} \bar g^{\mu\nu} \partial_\nu \sigma \right) &=& %
	- \frac{\partial V^{\mathrm{eff}}}{\partial\sigma} \, ,%
	 \label{eq:FE-sigma}%
\end{eqnarray}
with the effective energy--momentum tensor of the scalar graviton 
\begin{eqnarray} \label{Teff}
	\bar T^{\mathrm{eff}}_{\mu\nu} &=& {\bar {T_\sigma}}_{\mu\nu} + \bar g_{\mu\nu} \M2Pl\kappa_s e^{-2\bar w(\sigma) - \sigma} = 
	%
	\M2Pl \Upsilon^2 \left[ \partial_\mu \sigma \partial_\nu \sigma - \bar g_{\mu\nu} \left( \frac{1}{2}  (\partial\sigma)^2 - V^{\mathrm{eff}}(\sigma \right) \right] \, \nonumber , 
\end{eqnarray}
and the effective potential
\begin{equation} \label{Veff}
	V^{\mathrm{eff}}(\sigma) = V_\sigma(\sigma) + \frac{\kappa_s}{\Upsilon^2} e^{-2\bar w(\sigma) - \sigma} \, .
\end{equation}
Note that the integration constant $\kappa_s \ne 0$ makes the scalar graviton self-interacting even in case of the Lagrangian $V_\sigma(\sigma) \equiv 0$.

Equations (\ref{eq:Gmunu}), (\ref{eq:FE-sigma}) along with Eq.~(\ref{eq:FE-Z}) represent the full set of field equations (FE)
of the minimal SBR implementaion.

\section{Vacuum stationary spherically symmetric dark holes} \label{sec:dark-holes}

To study vacuum stationary spherically symmetric solutions of Eqs.~(\ref{eq:Gmunu}), (\ref{eq:FE-sigma}),
we define the line element in polar coordinates $r, \theta, \phi$ in the reciprocal gauge:
\begin{equation*}\label{eq:ds}
ds^2 \equiv \bar g_{\mu\nu}\, dx^\mu dx^\nu =  A(r) dt^2 - C(r) r^2 (\sin^2 \theta\, d\phi^2 + d\theta^2) - A^{-1}(r) dr^2 \, .
\end{equation*}
Eq.~(\ref{eq:Gmunu}) gives two independent equations for the metric coefficients $A$ and $C$~\cite{icppa-2024-proc}:
\begin{eqnarray}
	\frac{1}{2}	\frac{1}{r^2 C}\, \frac{d}{dr} \left( {r^2 C} \frac{dA}{dr} \right) = %
	- \Upsilon^2 V^{\mathrm{eff}}(\sigma) - \bar\Lambda   \label{FE-r-00} \, , \\
	\frac{1}{r^2 C^{1/2}}\, \frac{d}{dr} \left(r^2 \frac{dC^{1/2}}{dr} \right) %
	= %
	- \frac{1}{2} \Upsilon^2 {\left(\frac{d\sigma}{dr}\right)}^2 \label{FE-r-00-rr} \, .
\end{eqnarray}
Equation~(\ref{eq:FE-sigma}) gives:
\begin{equation}
	\frac{1}{r^2 C}\,  \frac{d}{dr}\left(r^2 C A \frac{d\sigma}{dr} \right) = %
	\frac{\partial V^{\mathrm{eff}}(\sigma)}{\partial\sigma} \label{FE-r-sigma} \, ,
\end{equation}
with $V^{\mathrm{eff}}(\sigma)$ defined by (\ref{Veff}).
Following Ref.~\cite{icppa-2024-proc}, we consider a simplified case with $V_\sigma \equiv 0$, $\bar\Lambda = 0$ and $\bar w(\sigma) = \bar w' \sigma$ with a constant $\bar w' \ne -1/2$.\footnote{%
We assume $|1+2\bar w'| \gg \Upsilon$ as for  $|1+2\bar w'| \sim \Upsilon$ the latter cannot be considered as a small parameter.
}
The effective potential (\ref{Veff}) takes a simplified form:
\begin{equation}\label{eq:Veff-simplified}
	V^{\mathrm{eff}}_\sigma(\sigma) = \frac{\kappa_s}{\Upsilon^2} e^{-(1+2\bar w')\sigma} \, .
\end{equation}
Under these assumptions, solutions of Eqs.~(\ref{FE-r-00})-(\ref{FE-r-sigma}) can be classified according to the sign of $\kappa_s$.

At $\kappa_s = 0$,
Eqs.~(\ref{FE-r-00})-(\ref{FE-r-sigma}) take Einstein -- massless scalar form, with the unique 
two-parametric asymptotically flat Fisher--Buchdahl--Janis--Newman--Winicour (FBJNW) solution~\cite{fisher,BJNW}\footnote{%
The solution was originally found in Ref.~\cite{fisher} in the astronomical ($C=1$) gauge.
Later, it was rediscovered~\cite{BJNW} using the reciprocal gauge admitting the explicit form given by Eqs.~(\ref{eq:Fisher-metric}), (\ref{eq:Fisher-sigma}).
}
featuring the metric with a naked singularity at $r = r_g$ 
and converging to Schwarzschild metric in the $\Upsilon \to 0$  limit:
\begin{equation} \label{eq:Fisher-metric}
  A(r) = \left(1 - \frac{r_g}{r}\right)^{1/\sqrt{1 + 2\Upsilon^2 \sigma_r^2}}     , \,\,
  C(r) = \left(1 - \frac{r_g}{r}\right)^{1 - 1/\sqrt{1 + 2\Upsilon^2 \sigma_r^2}}  \, ,
\end{equation}
\begin{equation} \label{eq:Fisher-sigma}
	\sigma(r) = \frac{\sigma_r}{\sqrt{1+2\Upsilon^2 \sigma_r^2}} \log\left(1-\frac{r_g}{r} \right) + \sigma_0 \, ,
\end{equation}
where $r_g$ is the central mass parameter and $\sigma_r$ parameterizes the strength of the scalar field $\sigma$ (defined up to an additive constant $\sigma_0$).
Limits on parameters of the FBJNW solution following from Solar system precision tests can be found, e.g., in \cite{Pirogov2011}.
In what follows, this solution is termed a {\it degenerate dark hole}.

In $\kappa_s > 0$ case, 
the scalar graviton mimics a dynamical DE, as studied in a cosmological context in \cite{Pirogov2018}.
Though it represents certain interest in a context of a localized stationary solution, we do not consider this case here.

At $\kappa_s < 0$, 
solutions of Eqs.~(\ref{FE-r-00})--(\ref{FE-r-sigma}) represent objects merging a central GR-like black hole (BH) and a peripheral scalar graviton halo~\cite{Pirogov2011}.
As shown in \cite{Pirogov2011},
at $\kappa_s < 0$ all stationary spherically symmetric vacuum solutions asymptotically converge to the {\it exceptional} three-parametric solution
featuring non-Minkowski asymptotic at $r\to\infty$,
with a property of gravitational confinement resulting in an asymptotically flat rotation curve.%
\footnote{%
Study~\cite{Pirogov2011} was performed in a context of a unimodular bimode gravity with the non-dynamical scalar density 
in the limit of scalar/tensor gravity coupling $|\Upsilon| \ll 1$. 
}

\paragraph*{The exceptional solution} 
presented in the exact form in Ref.~\cite{icppa-2024-proc} reads:\footnote{%
It may be noted that the exact exceptional solution coincides with the one obtained in a different context of Einstein--Maxwell--dilaton model in \cite{Sheykhi}.
}
\begin{equation} \label{eq:exceptional-metric}
	 A(r)  = \left(1 - \frac{r_g}{r}\right) \left(\frac{r}{r_h}\right)^{\frac{4\Upsilon^2}{(1+2\bar w')^2 + 2\Upsilon^2}}    , \,\,
	 C(r)  = \left(\frac{r}{r_c}\right)^{-\frac{4\Upsilon^2}{(1+2\bar w')^2 + 2\Upsilon^2}}   \, ,
\end{equation}
\begin{equation}\label{eq:exceptional-sigma}
	\sigma(r)   = \frac{2(1+2\bar w')}{(1+2\bar w')^2 + 2\Upsilon^2} \log\frac{r}{r_h}  \, , 
\end{equation}
where $r_g$ is the Schwarzschild radius of the central black hole,
the scalar profile parameter 
$r_h = \Upsilon \left[ - 2/((1+2\bar w')^2+2\Upsilon^2) \cdot 1/\kappa_s\right]^{1/2}$
and the $r_c$ parameter is fixed by the gauge-invariant geometric condition implying that at $r\to \infty$ the ratio of the measured circumference of a circle to its radius is $2\pi$:
$r_c = r_h \left[ 1 + {2\Upsilon^2}/(1+2\bar w')^2 \right]^{{(1+2\bar w')^2}/{2\Upsilon^2}}$
($r_c \simeq r_h e$ in the $\Upsilon \to 0$ limit).
The corresponding expressions for $Z^a$ fields are given in Appendix.
In the scalar graviton decoupling limit $\Upsilon \to 0$ metric (\ref{eq:exceptional-metric}) converges to the Schwarzschild metric.

At $\Upsilon \ne 0$, metric~(\ref{eq:exceptional-metric}) is not asymptotically flat and leads to gravitational confinement.
For a test particle orbiting the center at a constant radius $r$ one obtains the 
visible gauge-invariant rotation velocity~\cite{icppa-2024-proc},
\begin{equation}\label{eq:v-rot}
v_{\mathrm{rot}}(r) = \left[ \left( 1 + \frac{2\Upsilon^2}{(1+2\bar w')^2} \right) \, \frac{1}{2} \frac{r_g}{r-r_g} %
						  { \, + \,  \frac{2\Upsilon^2}{(1+2\bar w')^2} } \right]^{1/2},
\end{equation}
approaching a constant in the $r \to \infty$ limit:\footnote{%
$v_\infty$ grows as $1+2\bar w'$ approaches 0. The limiting case $\bar w' = -1/2$ requires special study.
Safely, our present consideration is limited to $\Upsilon \ll |1+2\bar w'|$ corresponding to $v_\infty \ll 1$.
}
\begin{equation}\label{eq:v-rot-asymptotic}
	v_{\infty} = \frac{\sqrt{2}\Upsilon}{|1+2\bar w'|} \, .
\end{equation}
This is similar to the phenomenon of asymptotically flat rotation curves, observed in galaxies~\cite{flat} 
and conventionally ascribed to presence of some DM with a properly chosen density profile~\cite{flat-review}.
As long as vacuum solutions are considered, they should be compared with DM dominated dwarf galaxies, 
with typical asymptotic rotation velocities $v_\infty \sim 30$~km/s~\cite{flat-review}.
For the scalar graviton halo, the asymptotic rotation velocity (\ref{eq:v-rot-asymptotic}) depends only on the Lagrangian parameters $\Upsilon$ and $\bar w'$.
If the scalar graviton is the only DM component, one may obtain a realistic $v_\infty$ estimate with $\Upsilon \sim 10^{-4}$ (assuming $|1+2\bar w'| \gg \Upsilon$).
However, if a more general case of larger galaxies with a lower relative DM content\footnote{%
As large galaxies are not dominated by DM, their rotation curves result from an interplay between distributions of baryonic and dark matter,
which makes the comparison with purely vacuum solutions inapplicable.
}	
is considered,
$\Upsilon$ values can be extended to $\sim 10^{-3}$ in order to obtain $v_\infty \sim 300$~km/s typical for large spiral galaxies~\cite{flat-review}.
This results in the scalar graviton's kinetic term scale $M_s = \Upsilon \MPl \sim 10^{14} \textendash 10^{15}$~GeV.\footnote{%
Is $M_s$ close to GUT scale by coincidence? 
}
On the other hand, metric~(\ref{eq:exceptional-metric}) with $\Upsilon$ values in the $10^{-4} \textendash 10^{-3}$ range 
may result in observationally unacceptable effects at finite $r$.
As will be shown below, the exceptional solution can be modified to improve the small distance behavior while preserving the asymptotic one.

\paragraph*{Modified exceptional solution.}
Before constructing a more general solution of Eqs.~(\ref{FE-r-00})-(\ref{FE-r-sigma}), its general properties\footnote{Not exposed in Ref.~\cite{icppa-2024-proc}.}
are to be recapitulated:

{\it (i) }
Setting $d\sigma(r)/dr = 0$ at some $r$  in (\ref{FE-r-sigma}) one finds:
\begin{equation}\label{eq:sigma-extremum}
	A(r) \frac{d^2\sigma}{dr^2} = - \frac{\kappa_s (1 + 2\bar w')}{\Upsilon^2} e^{-(1+2\bar w')\sigma} \, .
\end{equation}
Given $\kappa_s < 0$ and $A(r) > 0$, this implies that if $\sigma(r)$ has an extremum at some $r$, 
then it is a minimum for $1 + 2\bar w' > 0$ and a maximum for $1 + 2\bar w' < 0$, respectively.\footnote{Recall that we assumed $\bar w' = const$.}
Thus, $\sigma(r)$ either has a single extremum or no extrema at all. 
In what follows, we assume $1 + 2\bar w' > 0$ for definiteness.\footnote{The $1+2\bar w' < 0$ case can be considered by changing sign of $\sigma(r)$.}

{\it (ii)}
As shown in~\cite{Pirogov2011}, any stationary spherically symmetric solution asymptotically converges to the exceptional one at $r \to \infty$.

{\it (iii)}
Any solution is defined by 4 parameters.\footnote{%
 The solution is fully determined by $\kappa_s$ and $A$, $dA/dr$, $C$, $dC/dr$, $\sigma$, $d\sigma/dr$ values at some $r = r_0$.
 The potential (\ref{eq:Veff-simplified}) is invariant under simultaneous transformations: 
 $\sigma \to \sigma + \Delta\sigma$, 
 $\kappa_s \to \kappa_s e^{(1+2\bar w')\Delta\sigma}$, where $\Delta\sigma$ is an arbitrary constant shift.
 This reduces the number of free parameters by one.
 Fixing $A(r_0)$ by linear scaling of $t$ (and $r$, to preserve the reciprocal gauge) eliminates one more free parameter.
 Finally, $C(r_0)$ can be fixed by the asymptotic circumference/radius requirement.
 So far, only 4 independent parameters remain.
}

Let us note that $V^{\mathrm{eff}}(\sigma)$ vanishes at $\sigma \to +\infty$
and consider a modified solution with a $\sigma \to +\infty$ singularity at $r \to r_g$.
As in this limit $V_{\mathrm{eff}}(\sigma)$ can be neglected, 
the modified solution behaves like the degenerate one (\ref{eq:Fisher-metric}), (\ref{eq:Fisher-sigma}) at $r \to r_g$:
\begin{equation}\label{eq:modified-sigma-rg}
	\sigma(r) \sim \sigma_r \log\left(1 - \frac{r_g}{r}\right) + \sigma_0\, , 
\end{equation}
where $\sigma_r < 0$, $\sigma(r) > 0$ and $d\sigma(r)/dr < 0$.
On the other hand, at $r \to \infty$ the solution must converge to the exceptional one with $d\sigma/dr > 0$.
Thus, at some $r > r_g$ $\sigma(r)$ reaches its minimum.\footnote{In the special case $r_g = 0$ studied in \cite{Pirogov2011} the minimum is located at $r = 0$.}

It is convenient to define a ``Yukawa charge'' as a (distributed) source of the $\sigma$ field:
\begin{equation}\label{eq:Yukawa-def}
	Y_\sigma(r) \equiv 4\pi r^2 C A \frac{d\sigma}{dr} \, .
\end{equation}
The motivation is that for the degenerate solution (\ref{eq:Fisher-metric}), (\ref{eq:Fisher-sigma}) $Y_\sigma(r)$ does not depend on $r$ 
and can be indeed considered as a charge of the central source of $\sigma$ field:
\begin{equation}\label{eq:Yukawa-fisher}
	Y_{\sigma} = \frac{4\pi r_g \sigma_r}{\sqrt{1 + 2\Upsilon^2 \sigma_r^2}} \, .
\end{equation}
For the exceptional solution (\ref{eq:exceptional-metric}), (\ref{eq:exceptional-sigma}) $Y_\sigma(r)$ vanishes in the $r \to r_g$ limit.
The modified solution interpolating between the degenerate and exceptional ones has $Y_\sigma(r)|_{r \to r_g} < 0$.

The small distance behavior of the modified solution can be parameterized by the two free parameters, $r_g$ and $Y_\sigma(r)|_{r \to r_g}$.
The asymptotic behavior is characterized by the halo profile parameter $r_h$ and the intercept radius $r_{\mathrm{int}}$, 
i.e. the minumum $r$ where the modified $\sigma(r)$ crosses the asymptote given by Eq.~(\ref{eq:exceptional-sigma}) with the same $r_h$.
The $r_{\mathrm{int}}$ can be considered as the distance of the onset of gravitational confinement.

\begin{figure}[bp]
	\parbox{0.5\textwidth}{ \centering
		\includegraphics[width=0.5\textwidth]{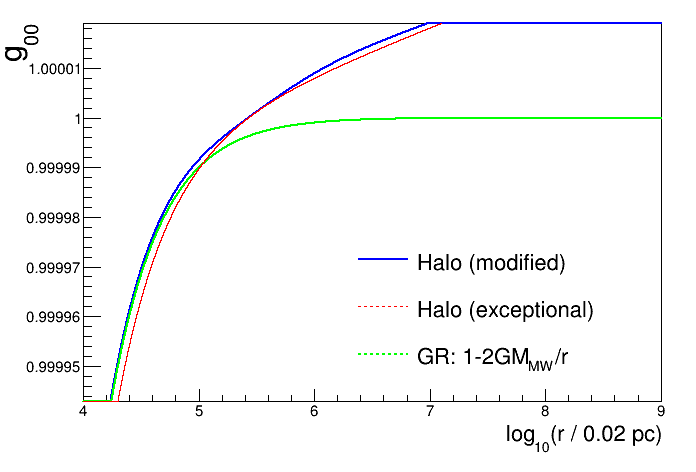}\\
		(a)
	}
	\parbox{0.5\textwidth}{	\centering
		\includegraphics[width=0.5\textwidth]{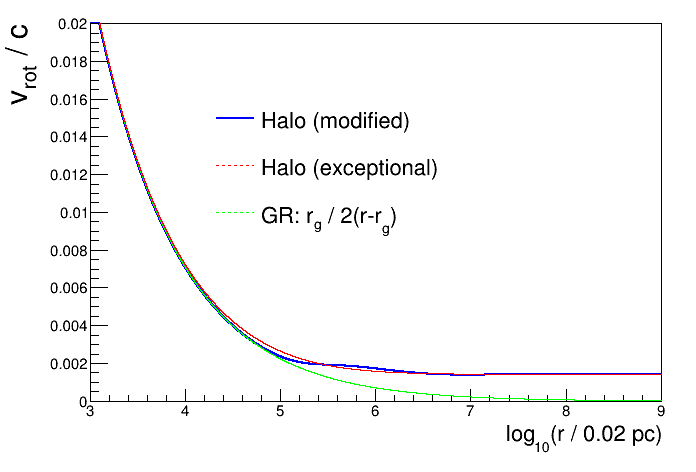}\\
		(c)
	}
	\parbox{0.5\textwidth}{	\centering
		\includegraphics[width=0.5\textwidth]{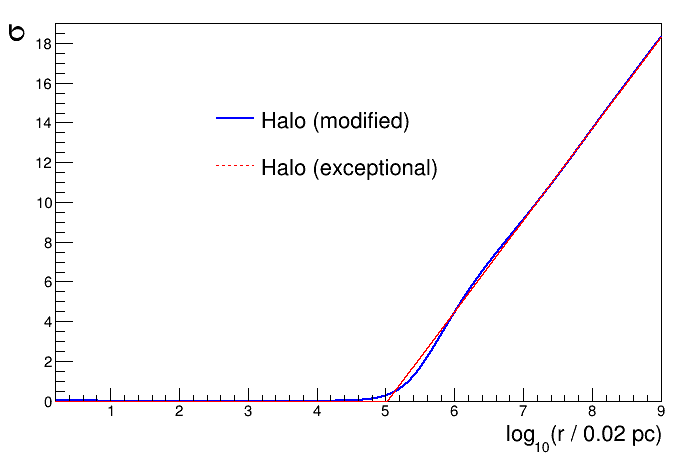}\\
		(b)
	}
	\parbox{0.5\textwidth}{	\centering
		\includegraphics[width=0.5\textwidth]{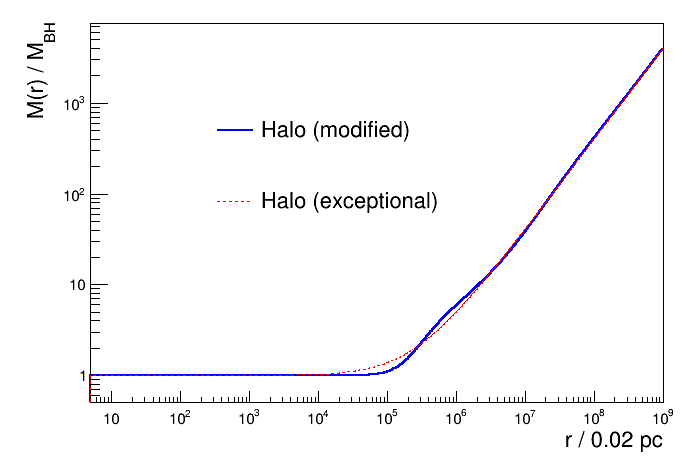}\\
		(d)
	}
	\caption{ \label{fig:exceptional-near-mod}
		The modified exceptional solution (blue curves):
		(a) $g_{00}$ metric component, (b) $\sigma(r)$, 
		(c) circular rotation velocity $v_{\mathrm{rot}}(r)/c$, 
		(d) the effective gravitating mass $M_g(r)$ normalized to the central black hole mass. 
		The same quantities are shown in red for the exceptional solution with the same $r_g$ and $r_h$.
		$g_{00}$ and  $v_{\mathrm{rot}}(r)$ for Schwarzschild metric with the same $r_g$ are shown in green.
		Central Yukawa charge and the intercept radius of the modified solution are 
		$Y_\sigma / r_g = -0.75$ and  $r_{\mathrm{int}} / 0.02 \mathrm{pc} = 10^5$, respectively
		(see the explanation of the parameters in the text).
	}
\end{figure}

\begin{figure}[htbp]
	\parbox{0.5\textwidth}{ \centering
		\includegraphics[width=0.5\textwidth]{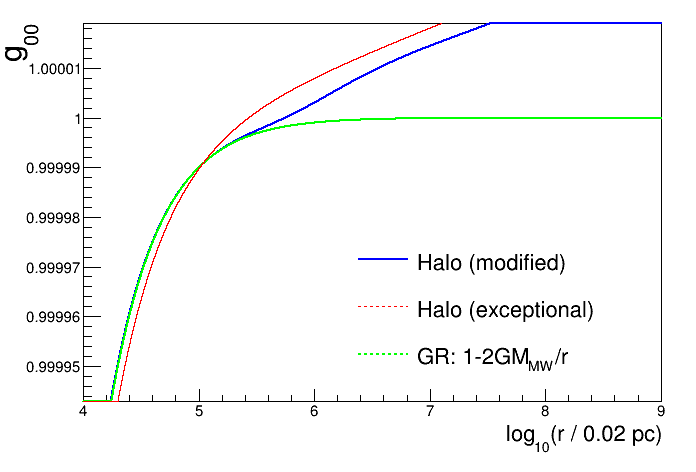}\\
		(a)
	}
	\parbox{0.5\textwidth}{	\centering
		\includegraphics[width=0.5\textwidth]{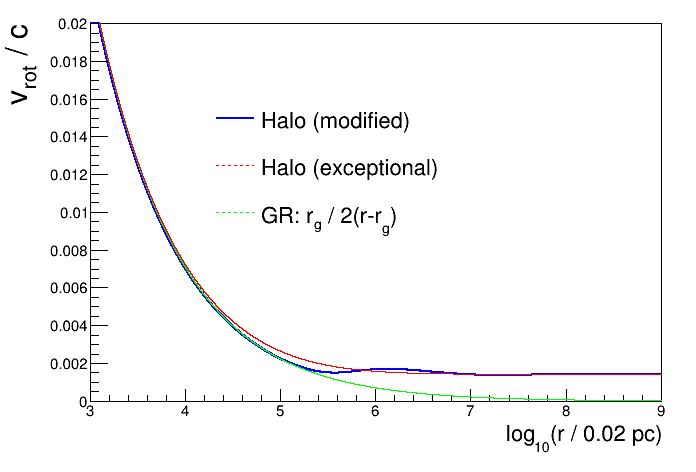}\\
		(c)
	}
	\parbox{0.5\textwidth}{	\centering
		\includegraphics[width=0.5\textwidth]{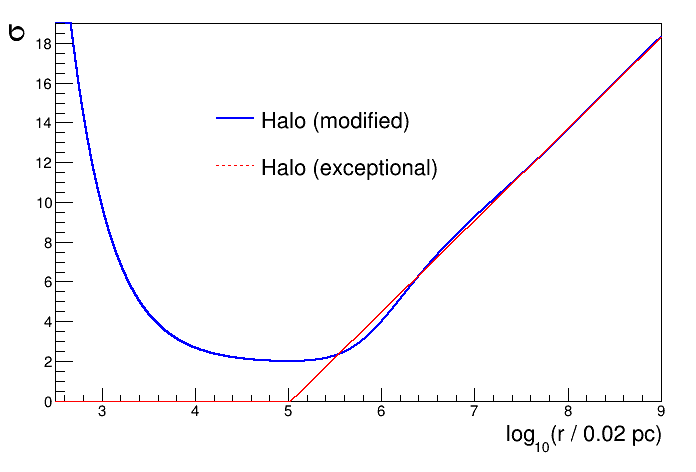}\\
		(b)
	}
	\parbox{0.5\textwidth}{	\centering
		\includegraphics[width=0.5\textwidth]{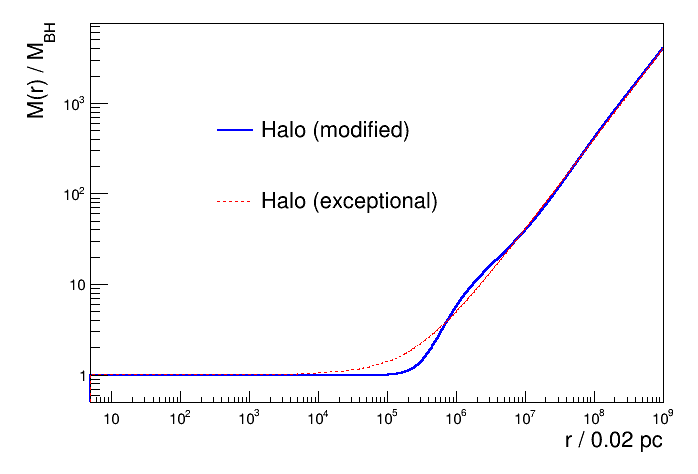}\\
		(d)
	}
	\caption{ \label{fig:exceptional-near-mod-1}
		Same as in Fig.~\ref{fig:exceptional-near-mod} with
		 $Y_\sigma / r_g = -0.997 \cdot 10^5$, $\log_{10} \left(r_{\mathrm{int}} / 0.02 \mathrm{pc} \right) = 5.43$.
	}
\end{figure}

\begin{figure}[htbp]
	\parbox{0.5\textwidth}{ \centering
		\includegraphics[width=0.5\textwidth]{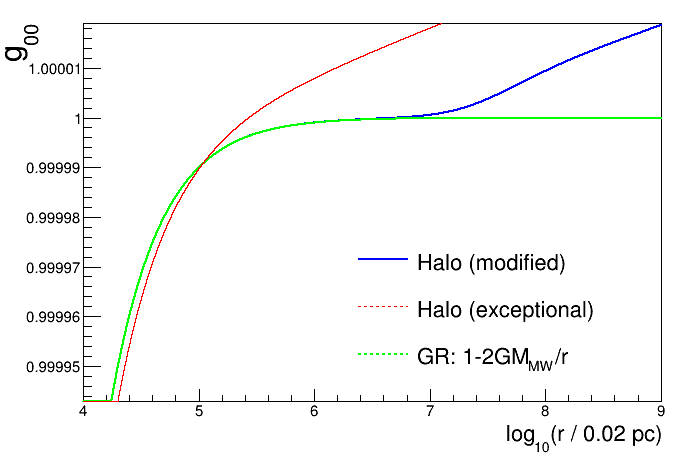}\\
		(a)
	}
	\parbox{0.5\textwidth}{	\centering
		\includegraphics[width=0.5\textwidth]{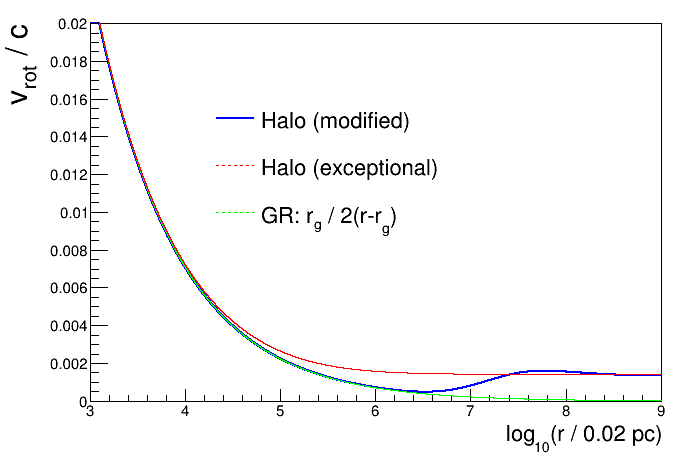}\\
		(c)
	}
	\parbox{0.5\textwidth}{	\centering
		\includegraphics[width=0.5\textwidth]{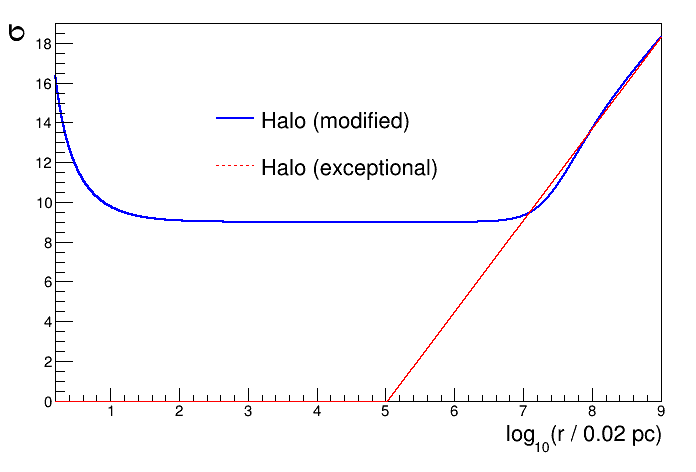}\\
		(b)
	}
	\parbox{0.5\textwidth}{	\centering
		\includegraphics[width=0.5\textwidth]{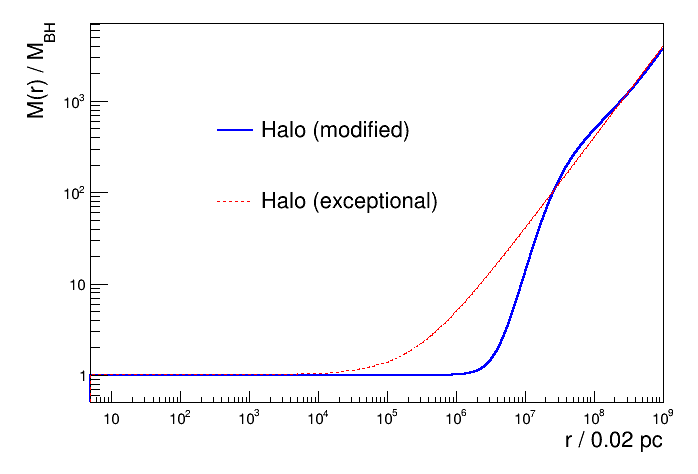}\\
		(d)
	}
	\caption{ \label{fig:exceptional-near-mod-2}
		Same as in Figs.~\ref{fig:exceptional-near-mod}, \ref{fig:exceptional-near-mod-1} with 
		$Y_\sigma / r_g = -0.93 \cdot 10^2$, $\log_{10} \left(r_{\mathrm{int}} / 0.02 \mathrm{pc} \right) = 6.95$.
	}
\end{figure}

\begin{figure}[htbp]
	\parbox{0.5\textwidth}{ \centering
		\includegraphics[width=0.5\textwidth]{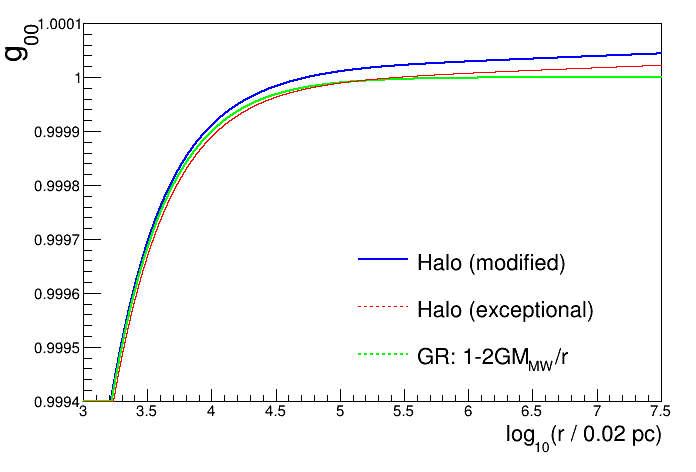}\\
		(a)
	}
	\parbox{0.5\textwidth}{	\centering
		\includegraphics[width=0.5\textwidth]{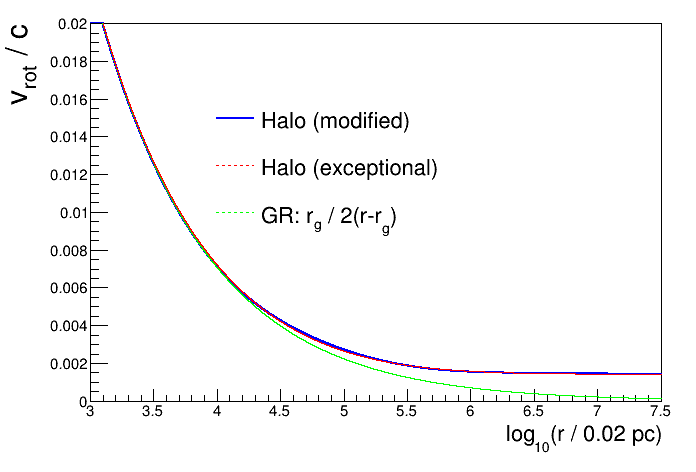}\\
		(c)
	}
	\parbox{0.5\textwidth}{	\centering
		\includegraphics[width=0.5\textwidth]{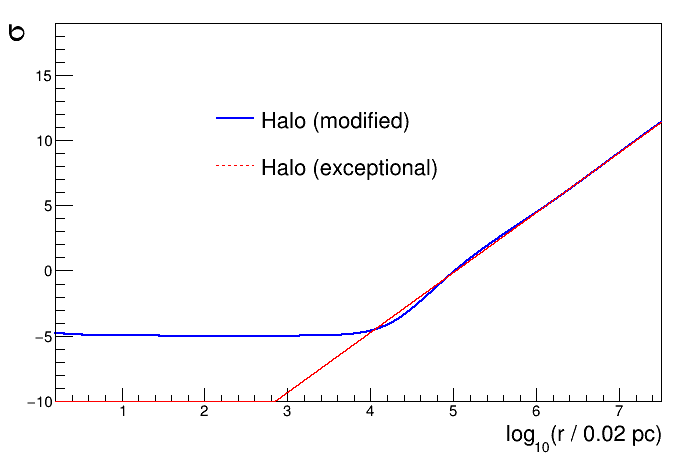}\\
		(b)
	}
	\parbox{0.5\textwidth}{	\centering
		\includegraphics[width=0.5\textwidth]{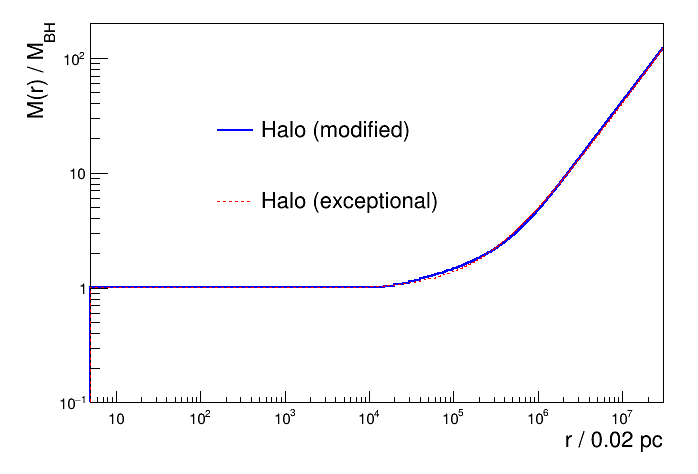}\\
		(d)
	}
	\caption{ \label{fig:exceptional-near-mod-3}
		Same as in Figs.~\ref{fig:exceptional-near-mod}-\ref{fig:exceptional-near-mod-2} with 
		$Y_\sigma / r_g = -3$, $\log_{10} \left(r_{\mathrm{int}} / 0.02 \mathrm{pc} \right) = 3.91$.
	}
\end{figure}

Given qualitative considerations above, we constructed modified solutions for representative set of free parameters
by integrating Eqs.~(\ref{FE-r-00})-(\ref{FE-r-sigma}) numerically.
The examples are shown in Figs.~\ref{fig:exceptional-near-mod}-\ref{fig:exceptional-near-mod-3}.
The plots show the metric component $g_{00}$, the strength of the scalar graviton field $\sigma$, 
the effective gravitating mass\footnote{Defined as 
$M_g = 2\M2Pl \int R^0_0 \sqrt{-\bar g}\, d^3x$, following Ref.~\cite{icppa-2024-proc}.
In GR, this definition coincides with Tolman's mass-energy of a spatially localized system~\cite{Tolman}.
Note that $M_g$ is gauge dependent.
}
$M_g$ normalized to the mass of the central black hole $M_{\mathrm{BH}} = r_g c^2/2G_N$, and 
the corresponding circular rotation velocity $v_{\mathrm{rot}}$ as functions of $r$.
On all plots, the Lagrangian parameters are fixed under $\Upsilon = 10^{-3}$, $\bar w' = 0$.%
\footnote{$\bar w' = 0$ corresponds to the spontaneously broken GR.
$\Upsilon = 10^{-3}$ corresponding to $v_\infty \sim 300$~km/s is chosen to amplify expected deviations from GR-like behavior.
However, in reality these deviations are smaller for $\Upsilon \sim 10^{-4}$.
}
The central black hole mass is chosen as $M_{\mathrm{BH}} = 2 \cdot 10^{11} M_{\odot}$, with 
the corresponding Schwarzschild radius $r_g \simeq 0.02$~pc.
The halo profile parameter is $r_h = 10^5 r_g \simeq 2$~kpc.
Variable parameters are the central Yukawa charge $Y_\sigma$ and the intercept radius $r_{\mathrm{int}}$.
The $g_{00}$ and $v_{\mathrm{rot}}$ plots demonstrate relatively slight deviation of the modified solution 
from Schwarzschild behavior up to $r \sim r_{\mathrm{int}}$, 
compared to sizable deviation of the exceptional solution.
This allows to satisfy possible observational constraints at short distances $r < r_{\mathrm{int}}$.
Unfortunately, this makes the predictions for dark holes less unique at small distances.

On the other hand, the family of modified solutions possesses essentially non-flat common asymptotic 
given by the exceptional solution (\ref{eq:exceptional-metric}), (\ref{eq:exceptional-sigma}).
This limits the applicability of the considered solutions for modeling realistic finite DM haloes.
Possible paths to resolve this problem are discussed below.

\section{Discussion}

Can dark holes be modified to model realistic cosmic objects with finite DM haloes?
The desirable features are Minkowski (or de Sitter) asymptotic at $r\to \infty$ and, simultaneously, 
an approximately flat rotation curve at intermediate $r$, up to some cutoff distance $r_{\mathrm{cut}}$.
For the stationary spherically symmetric configuration,
an immediate possibility is 
to introduce previously neglected effective CC approximating the DE potential term in Eq.~(\ref{eq:Ltensor}), $\bar V_{{\ae}} = \M2Pl \bar\Lambda$.
From Eqs.~(\ref{FE-r-00})-(\ref{FE-r-sigma}) one obtains for the $A(r)$ metric coefficient, up to terms linear in $\Upsilon^2$, $\bar\Lambda$ and $r_g$:
\begin{equation}\label{eq:A-with-Lambda}
A(r) = 1 - \frac{r_g}{r} + \frac{4\Upsilon^2}{(1+2\bar w')^2 } \log\frac{r}{r_h} - \frac{\bar\Lambda}{3} r^2 + \dots \, ,
\end{equation}
which for $\bar\Lambda > 0$ implies the cutoff at $r_{\mathrm{cut}} \simeq \sqrt{6/(1+2\bar w')} \Upsilon {\bar\Lambda}^{-1/2}$.
With ${\bar \Lambda}^{-1/2}$ of an order of the cosmological horizon scale and $\Upsilon \sim 10^{-4} \textendash 10^{-3}$,
one obtains $r_{\mathrm{cut}} \sim 1 \textendash 10$~Mpc, a scale relevant to galaxy clusters rather than to individual galaxies.
A more general approximation for the DE potential can be considered as well.

\begin{figure}[htbp]
	\parbox{0.5\textwidth}{ \centering
		\includegraphics[width=0.5\textwidth]{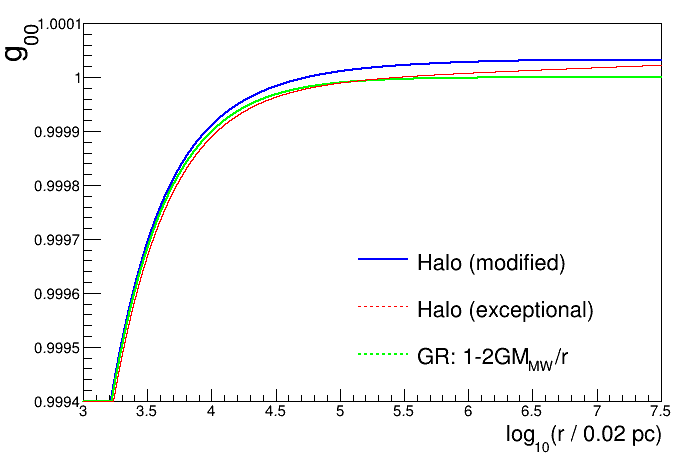}\\
		(a)
	}
	\parbox{0.5\textwidth}{	\centering
		\includegraphics[width=0.5\textwidth]{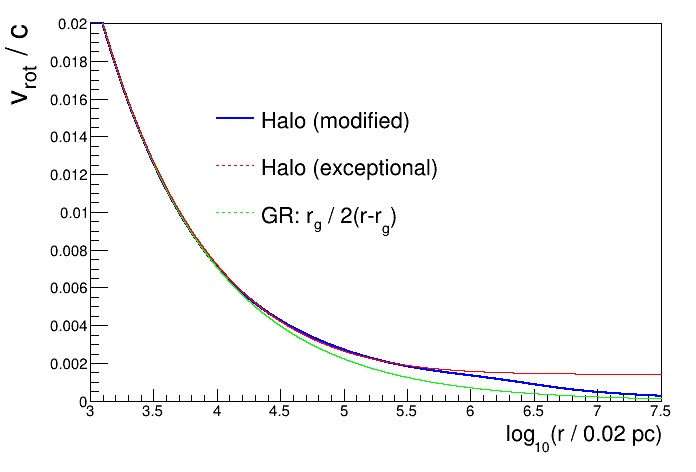}\\
		(c)
	}
	\parbox{0.5\textwidth}{	\centering
		\includegraphics[width=0.5\textwidth]{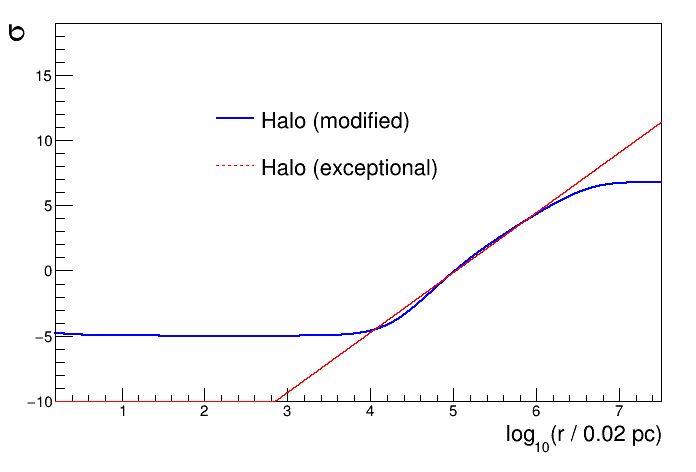}\\
		(b)
	}
	\parbox{0.5\textwidth}{	\centering
		\includegraphics[width=0.5\textwidth]{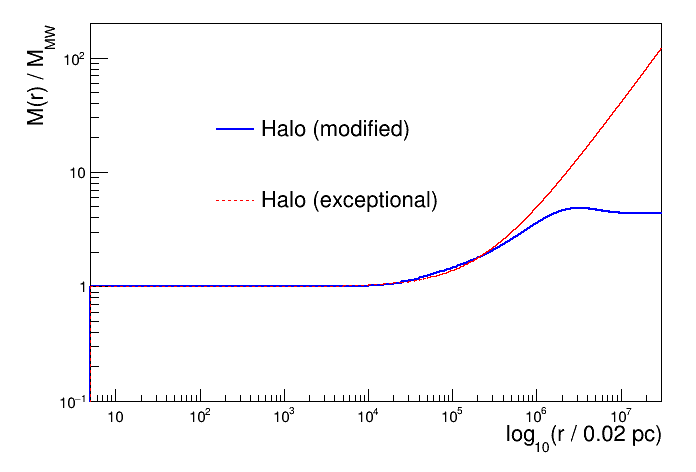}\\
		(d)
	}
	\caption{ \label{fig:exceptional-near-mod-3-mass-1}
		Same as in Fig.~\ref{fig:exceptional-near-mod-3}, with the additional quadratic $V_\sigma$ potential defined by Eq.~\ref{eq:Vsigma-quadratic}.
		The $V_\sigma$ parameters are $\sigma_0 = 7.55$ and $1/m = 38.8$ kpc, $V_0$ is fixed by the Minkowski asymptotic requirement. 		
		The other parameters of the solution are as in Fig.~\ref{fig:exceptional-near-mod-3}.
		Note the cutoff of the gravitational confinement regime at $r_{\mathrm{cut}} \simeq 20$~kpc.
		The effective gravitating mass (d) of the system is finite, contrary to the divergent mass in the exceptional solution.
	}
\end{figure}

\begin{figure}[htbp]
	\parbox{0.5\textwidth}{ \centering
		\includegraphics[width=0.5\textwidth]{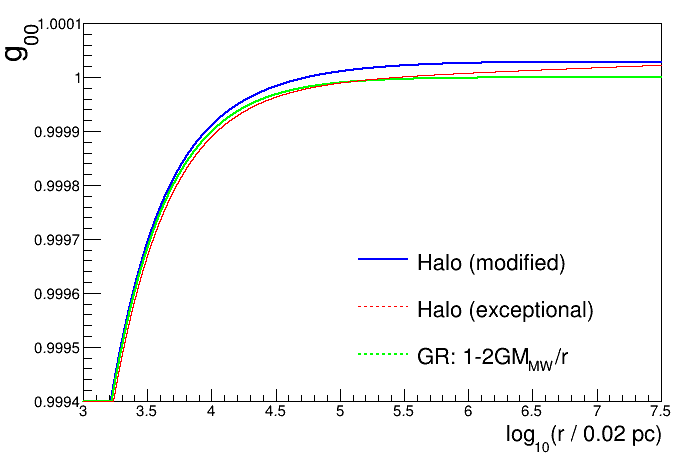}\\
		(a)
	}
	\parbox{0.5\textwidth}{	\centering
		\includegraphics[width=0.5\textwidth]{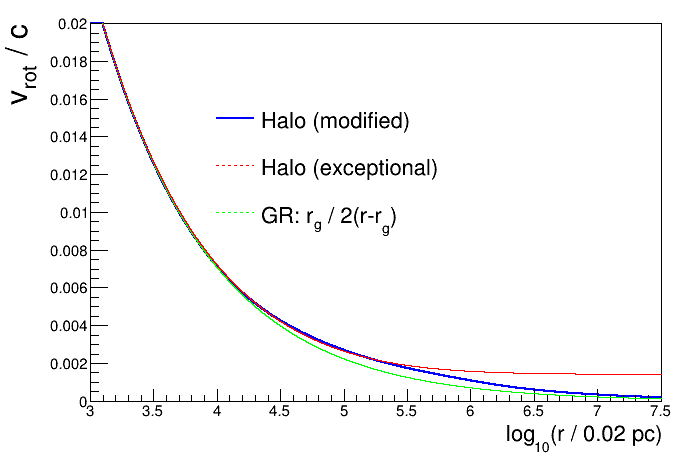}\\
		(c)
	}
	\parbox{0.5\textwidth}{	\centering
		\includegraphics[width=0.5\textwidth]{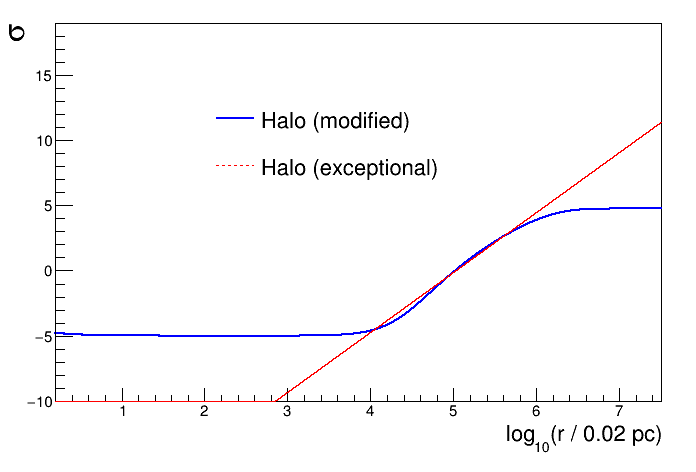}\\
		(b)
	}
	\parbox{0.5\textwidth}{	\centering
		\includegraphics[width=0.5\textwidth]{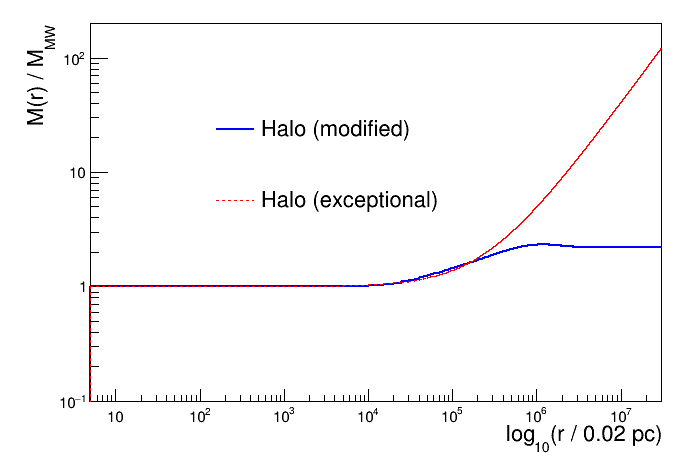}\\
		(d)
	}
	\caption{ \label{fig:exceptional-near-mod-3-mass}
		Same as in Fig.~\ref{fig:exceptional-near-mod-3-mass-1} with 
		$\sigma_0 = 5.55$, $1/m = 14.4$ kpc. 
		The other parameters are as in Fig.~\ref{fig:exceptional-near-mod-3}.
		Note the cutoff of the gravitational confinement regime at $r_{\mathrm{cut}} \simeq 8$~kpc.
		The effective gravitating mass (d) of the system is finite, contrary to the divergent mass in the exceptional solution.
	}
\end{figure}

Another possibility is a massive scalar graviton.
The minimal Lagrangian (\ref{eq:L})-(\ref{eq:Lscalar}) can be modified by 
introducing the potential $V_\sigma$ in (\ref{Veff}), 
such that the effective potential $V^{\mathrm{eff}}(\sigma)$ has a local minimum at some finite $\sigma = \sigma_{\mathrm{min}}$
but retains the exponential asymptotic at $\sigma \to -\infty$.%
\footnote{$V_\sigma \ne \mathrm{const}$ excplicitly violates the invariance under constant $\sigma$ shifts.}
For illustration purposes, let us choose the quadratic potential, 
\begin{equation}\label{eq:Vsigma-quadratic}
	V_\sigma(\sigma) = V_0 + \frac{m^2}{2} (\sigma - \sigma_0)^2  \, .
\end{equation}
At $\kappa_s < 0$, $V^{\mathrm{eff}}(\sigma)$ possesses a local minimum under the condition:
\begin{equation}
	m^2 > \left(-\frac{\kappa_s}{\Upsilon^2}\right) (1+2\bar w')^2 e^{1 - (1+2\bar w')\sigma_0} \, .
\end{equation}
The presence of the minimum allows, particularly, for a trivial stationary solution with $\sigma = \sigma_{\mathrm{min}}$ and the spacetime with a constant spatial curvature.
Solutions behaving like the (modified) exceptional solution at finite distances and asymptotically converging to the trivial one were studied numerically
for representative values of the $m$ and $\sigma_0$ parameters.
For definiteness, the $V_0$ parameter was adjusted to obtain asymptotically Minkowski space.\footnote{Minkowski asymptotic allows to consistently define a finite effective mass of the system.}
The examples are shown in Figs.~\ref{fig:exceptional-near-mod-3-mass-1}, \ref{fig:exceptional-near-mod-3-mass}.

Further Lagrangian modifications may include a non-minimal kinetic term of the scalar graviton and 
an account for a dynamical DE via $V_{{\ae}}(\bar{{\ae}}^\mu_\nu)$ making the tensor graviton massive~\cite{Pirogov2021}.
A deeper understanding of the underlying theory requires to study an influence of the ordinary matter on vacuum solutions, an aspect neglected so far.
Additional DM components (possibly coupled to the scalar graviton) can be considered as well.

More radical modifications concerning the definition of the effective metric 
(\ref{eq:gbar-ash}) include a non-linear or matter dependent Weyl form-factor $\bar w(\sigma)$.\footnote{%
E.g., $\bar w'(\sigma) \to -1/2$ at $\sigma \to \pm\infty$ and $-1/2 < \bar w'(\sigma) < 0$ at finite $\sigma$
may result in (anti)de Sitter asymptotic at $r\to \infty$, a GR-like solution at small distances and a gravitational confinement at intermediate $r$.
}
An important special case is $\bar w = -\sigma/2$, corresponding to the spontaneously broken Weyl-transverse relativity 
opening a possibility to resolve the CC problem~\cite{Pirogov2021}.

The proposed modifications require further extensive studies.

\section{Conclusion}

As an entrance point to a more general problem of merging a modified gravity, 
DM and DE through the multiscalar-metric gravity with spontaneously broken relativity (SBR), 
we considered basic vacuum stationary spherically symmetric solutions arising in the SBR framework -- {\it dark holes}, 
which merge a GR-like central black hole with an extending scalar graviton halo.
The manifestation of the latter is the effect of asymptotically flat rotation curves,
similar to the one observed in galaxies and generally attributed to the presence of some conventional DM.
Possibilities of further modification of basic dark holes are discussed, with a prospect to describe realistic astrophysical objects.
The dark holes may be considered as a distinctive feature of the multiscalar-metric paradigm.

\section*{Appendix}

The $Z^a$ fields corresponding to the exceptional solution (\ref{eq:exceptional-metric}), (\ref{eq:exceptional-sigma}) are conveniently 
expressed in internal spherical coordinates (up to global Lorentz transformations and shifts in the internal $\{Z^a\}$ space):
\begin{equation}
	Z^0 = C_0 t\, ,  \, Z^\theta = \theta\, , \, Z^\phi = \phi\, , \nonumber 
\end{equation}
\begin{equation}
	Z^r = r_h \left[\frac{3}{C_0} \cdot \frac{(1+2\bar w')^2+2\Upsilon^2}{(1+2\bar w')(1 + 6\bar w') + 2\Upsilon^2} \left(\frac{r_c}{r_h}\right)^{\frac{4\Upsilon^2}{(1+2\bar w')^2 + 2\Upsilon^2}} \left(\frac{r}{r_h}\right)^{\frac{(1+2\bar w')(1+6\bar w') + 2\Upsilon^2}{(1+2\bar w')^2 + 2\Upsilon^2}} \hspace*{-2ex} + C_1 \right]^{1/3} \, ,
\end{equation}
where $C_0$ and $C_1$ are arbitrary constants, and the parameters $r_h$, $r_c$ are defined in Section~\ref{sec:dark-holes}.

Note that $Z^a$ (and, consequently, $\sigma$) do not depend on the central mass parameter $r_g$. 

\vspace*{5ex}

\paragraph*{Acknowledgements.}
The authors are grateful to V.~V.\ Bryzgalov for discussions.

\end{document}